\documentclass[aps,prb,twocolumn,citeautoscript]{revtex4}
\usepackage{graphicx}
\usepackage{amsmath}
\usepackage{epsfig}

\newcommand{\ba}{\begin{eqnarray}}
\newcommand{\ea}{\end{eqnarray}}

\newcommand{\eaa}{\end{eqnarray*}}
\newcommand{\w}{\omega}
\begin{document}

\title{ Low Temperature Transport Properties of Strongly Interacting Systems \\
- Thermal Conductivity of Spin-$1/2$ Chains\\
{\small Dedicated to Nando Mancini on the occasion of his 60th birthday} }

\author{N. Andrei}
 \affiliation{Center for Materials Theory,
Rutgers University, Piscataway, NJ 08854--8019}
\author{E. Shimshoni}
\affiliation{Department of Mathematics--Physics, 
University of Haifa at Oranim, Tivon 36006, Israel}

 \author{ A. Rosch}
\affiliation{Institut f\"ur Theorie der Kondensierten Materie,
Universit\"at Karlsruhe, D-76128 Karlsruhe, Germany}

\begin{abstract}

 We outline a general approach to the computation of transport
properties of interacting systems at low temperetures and
frequencies. We show that if the fixed point and the irrelevant
operators around it are known, then by studying the structure of the
softly violated conserved currents chracterizing the fixed point one
may set up an effective calculation in terms of a memory matrix
formalism. We apply this approach to the computation of thermal
conductivity of spin chains embedded in a matter matrix and
interacting with its phonons.  The results are found to be in very good 
agreement with experiment.

\end{abstract}

\maketitle

The study of transport properties of strongly interacting systems has
been of great theortical and experimental interest for a long time. We
shall concentrate in this contribution on developing an effective
approach to the problem when the system is not far from its fixed
point; in other words, when it is probed at low temperatures and at
low frequencies. Subsequently we shall carry it out in detail for a
system of spin 1/2 chains coupled to phonons.

Our approach consists of the following elements:

{\bf 1.} Identify the fixed point of the hamiltonian which describes the
   system, as well as the irrelevant operators around it.

 The fixed point itself is typically insufficient to describe low
 energy transport properties; it is scale invariant and
 translationally invariant and thus unable to degrade a current,
 leading to infinite conductivity. To obtain finite conductivity one
 needs to take into account terms which break translational
 invariance and, more generally, violate the conservation laws associated with the
 fixed point.

{\bf 2.}  Study the (weakly violated) conserved charges around the fixed
    point hamiltonian.
    
    A fixed point $H^*$ is scale invariant and often has several
    conserved quantities $P$, $[P, H^*] =0$, associated with it. In
    1-d, for example, if the fixed point is conformally invariant it
    has an infinite number of conserved quantities. When the
    irrelevant operators around the fixed point are taken into
    account, most of these quantities no longer commute with the low
    energy hamiltonian. The conservation of most of these quantites is
    strongly violated, but some may be only weakly violated and then
    significantly influence the low energy dynamics of the system.

 Typically the current whose correlations determining the transport
  properties under considerations will be among those almost conserved
  charges, or ``protected'' by them in the following sense.  When a
  system possesses some conserved quantities $P$, these may
  ``protect'' the current $J$ from degrading (this occurs when the
  cross-susceptibility $\chi_{JP} \neq 0$) leading to a pure
  (i.e. $\delta(\omega)$) Drude peak and infinite d.c.  conductivity.
  When the conservation of the pseudo-momenta $P$ is  softly
  violated they will, instead, lead to very long time tails in the
  decay of the current $J$. This occurs since states with a finite
  pseudo-momentum $P$ typically carry also a finite current $J$ since
  $\chi_{JP} \neq 0$.  The component of the current ``parallel'' to
  $P$, $J_{\parallel P} =(\chi_{PJ} /\chi_{PP}) P$ will therefore
  decay slowly.  The presence of such approximately
  conserved quantities leads then to a natural hydrodynamic
  description of the system where a separation of fast and slowly
  decaying modes takes place and a consistent scheme of calculation of
  the slow mode conductivities can be carried out in terms of matrix
  of decay rates of these modes.

Let us explain and illustrate these ideas in more detail. We begin by
 arguing that given a conserved charge $P$, it will ``protect'' $J$ if
 $\chi_{PJ} \neq 0$. Indeed, imagine preparing at $t = 0$ a state
 carrying a current $\langle J \rangle$.  Then necessarily that state
 will also have a non vanishing $\langle P \rangle$,
\begin{eqnarray}
\langle P \rangle = 
\frac{\chi_{P  J}}{\chi_{JJ}}  \langle J \rangle. \nonumber
\end{eqnarray}

\begin{center}
\includegraphics[width=0.50 \linewidth]{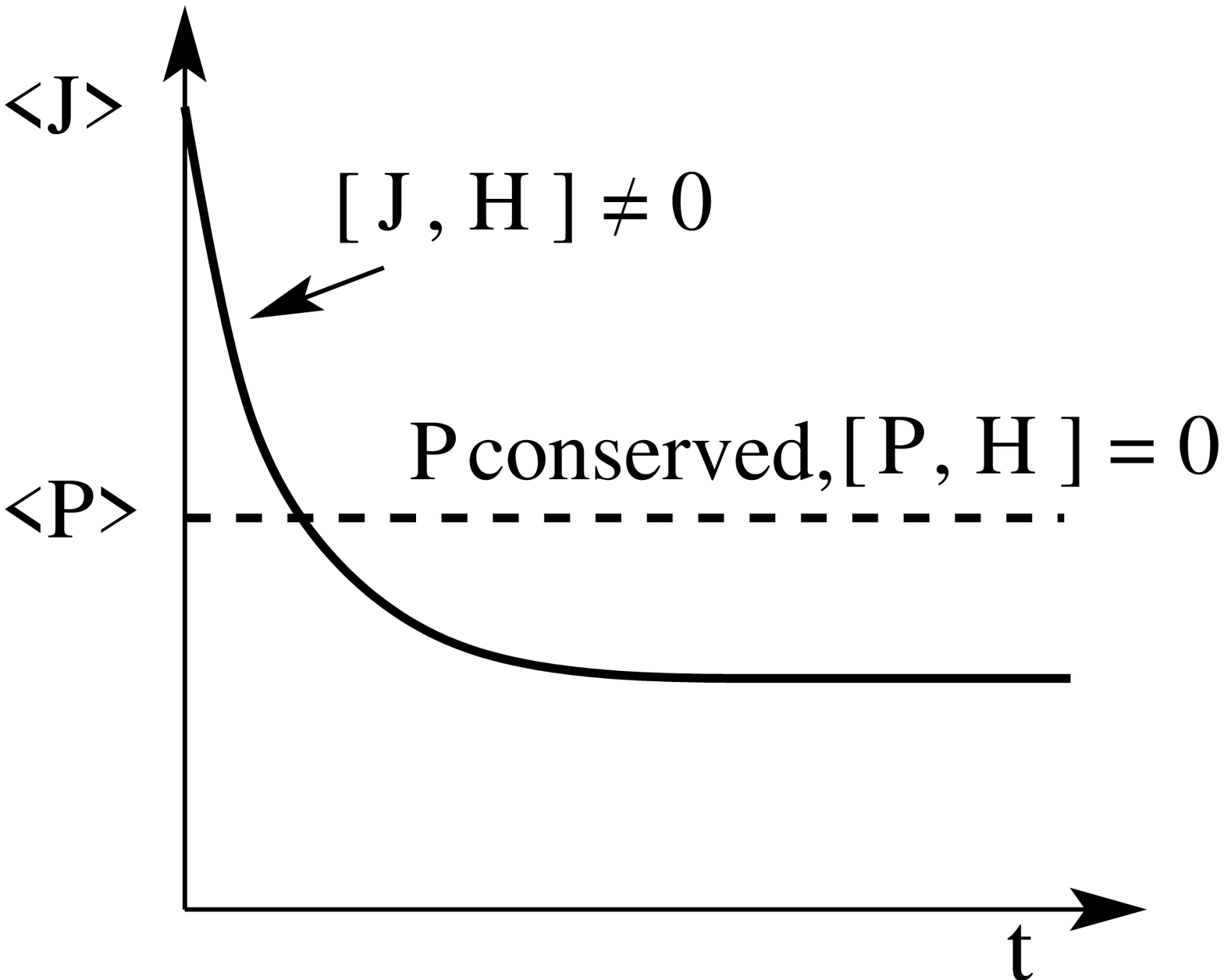}
\end{center}
\centerline{FIG. 1}
\bigskip

In the limit  $t \to \infty$  since $P$ is conserved its expectation value 
will not have changed, and it will in its turn induce a non vanishing 
expectation value for $J$,
\begin{eqnarray*}
 \lim_{t \to \infty}\langle J \rangle = 
\frac{\chi_{JP}}{\chi_{P P}}\langle P \rangle = 
\frac{\chi_{ JP}^2}{\chi_{P P} \chi_{JJ}} \langle  J(t=0) \rangle 
\end{eqnarray*}
(see Fig. 1).  Since the current tends asymptotically to a constant
value we find that the conductivity will have a $\delta(\omega)$-Drude
peak containing a fraction $\frac{\chi_{JP}^2}{\chi_{P P} \chi{JJ}}$
of the total weight; in other words, a Drude weight $D=\frac{1}{2}
\frac{\chi_{ JP}^2} {\chi_{ PP}}$. An immediate consequence is that
integrable models,  having  infinitely many conserved
quantities,  will typically also have an infinite dc conductivity.

If the charge  $P$ is not conserved but  slowly decaying it will induce
  slow (long-time) decay in  $J$ (see Fig. 2).
  
\begin{center}
\includegraphics[width=0.5 \linewidth]{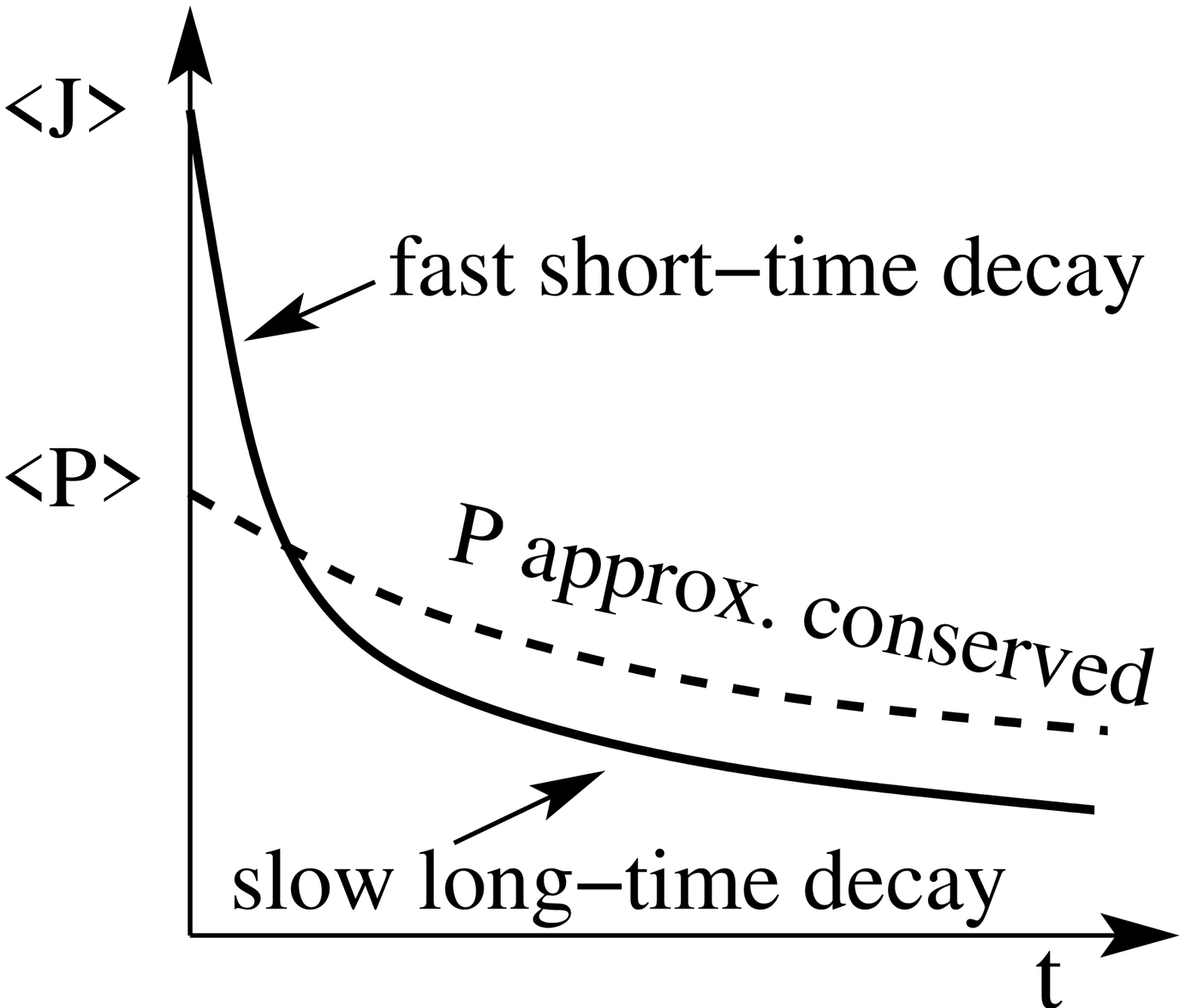}
\includegraphics[width=0.4 \linewidth]{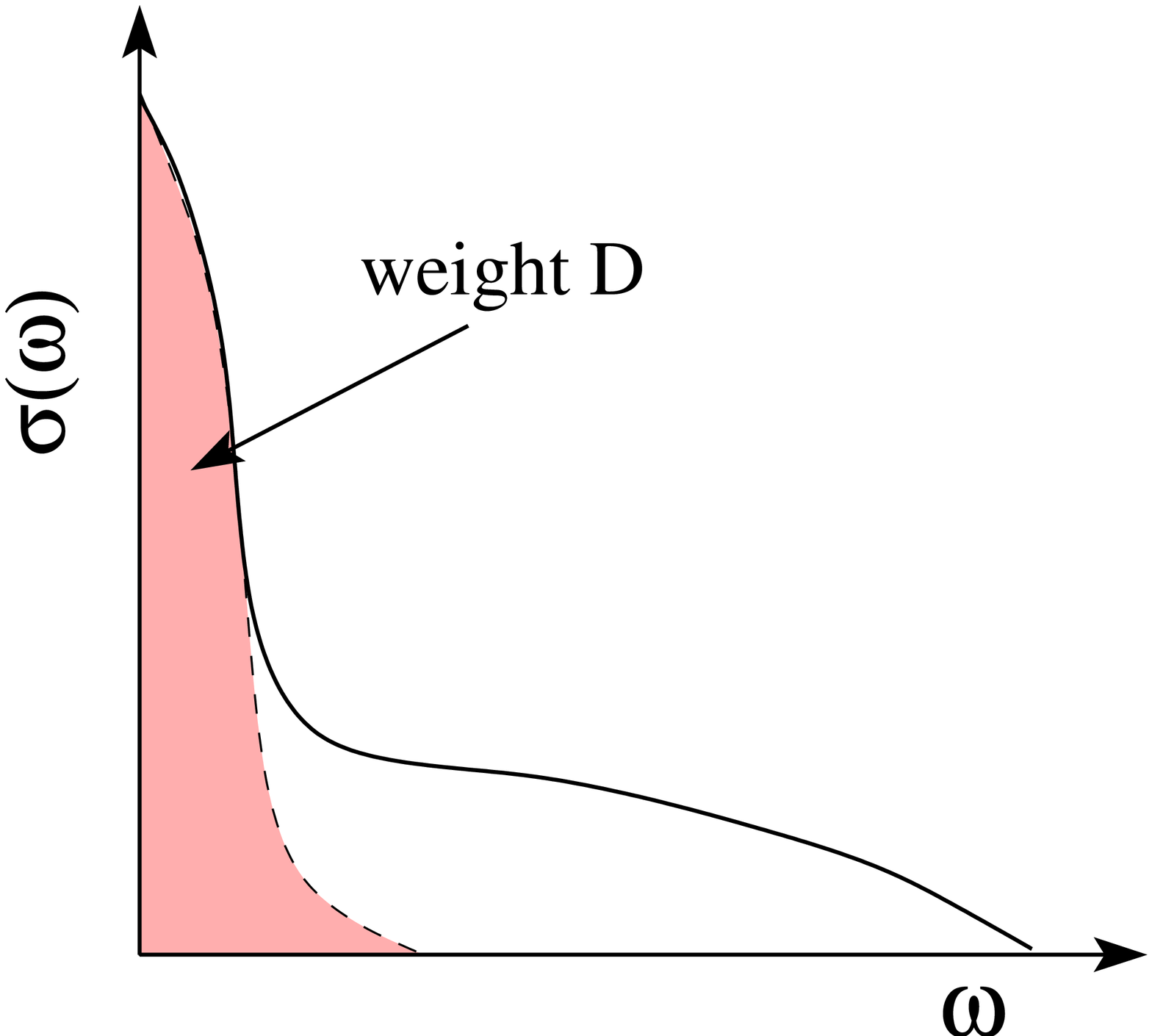}
\end{center}
\centerline{FIG. 2}
\bigskip

This slow decay will show up as a peak in $\sigma(\omega)$ and the
decay-rate $\Gamma_P$ of $P$ will determine the width of the peak
  and the dc value $\sigma(0)\approx D/\Gamma_P$ of the optical
  conductivity, the area under the peak being again, $D=\frac{1}{2}
\frac{\chi_{JP}^2}{\chi_{PP}} $.

{\bf 3.} Having identified the slowly decaying charges of the effective model
  describing the low energy properties of the system we can compute
  the low temperature dc transport using a method that separates fast
  and slow modes, incorporating the former in the dynamics of the
  latter. Such a method is the memory matrix approach which can be
  used very efficiently and controllably when combined with the RG
  considerations outlined above.

\bigskip

Let us apply this approach to a system of spin chains embedded in some 
3-d lattice and interacting with its phonons. Such systems (including,
in particular, various compounds of SrCuO)
have been recently studied in detail by, e.g., Sologubenko {\it et al}\cite{solo}.
The authors have measured the heat conductivity along the three main axes
of the sample. They observed that while conductivities along the $a$ and $c$ 
axes almost coincide, the conductivity along $b$, the axis along which
the spin chains lie, has an enhancement which they interpret as being
due to contribution of the spin degrees of freedom
(Fig. 3, for example, presents data from a corresponding
measurement in Sr$_2$CuO$_3$). In the temperature range
$60\,$K $\leq T\leq 200\,$K, they gave the fit:
\begin{eqnarray*}
 \kappa_s(T) &\sim& \exp{(T^*/T)}\; , \\
 T^* &\approx& 0.42 \Theta_D 
\end{eqnarray*}
where $\kappa_s$ is the spin contribution obtained after subtracting the 
phonon background from $\kappa_b$.
The relevant energy scales of the system are: The Debye temperature
characterizing the phonons - $\Theta_D \sim \; 400\,  \;$K and the 
spinon interaction scale $J/k_B\sim 2600\,$K  characterizing the spin chain.

\bigskip
\begin{figure}[h]
\includegraphics[width=3.3in]{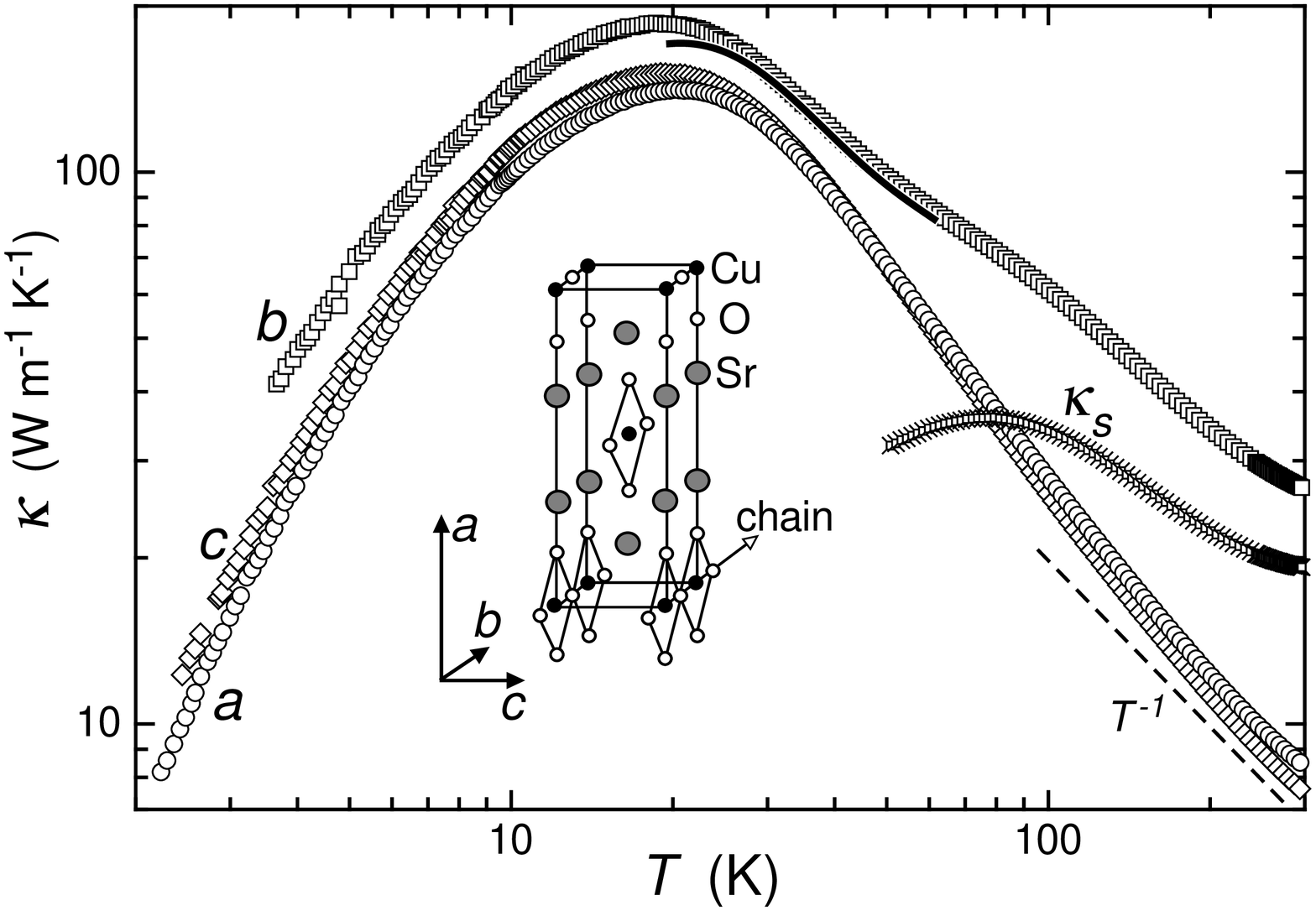}
\end{figure}
\centerline{FIG. 3}
\bigskip

One immediate question that arises is why is $\kappa_s$ actually
determined by the lower scale $\Theta_D$. Furthermore, by plotting the
above data of $\ln \kappa_s$ and of the pure phonon contribution 
$\ln \kappa_{a,c}$ vs. $1/T$, one observes that the slope of
the latter is larger by a factor of 2. Again, why?

We shall find that a rather subtle interplay of (approximate)
conservation laws and quantum dynamics underlies  the
experimentally observed heat conductivity, and the approach 
outlined above is necessary to fully account
for it.

\bigskip

We begin by discussing the low energy effective hamiltonian. First consider 
a single spin chain,

\begin{eqnarray*}
H_s=\frac{1}{2}\sum_{i,j=1}^N  J_{ij}
\left(S_i^+S_{j}^- + S_i^-S_{j}^+\right)
+\sum_{i,j=1}^N J_{ij}^z S_i^zS_{j}^z \\
\end{eqnarray*}

 As is well known\cite{affleck}, for  spin chains with short range interactions
 the fixed point hamiltonian is the Luttinger liquid,

\begin{eqnarray*}
 H_{LL}=-i (Ja)\int dx (\psi^\dagger_R\partial_x \psi_R -
\psi^\dagger _L\partial_x \psi_L) 
+ J_z \int dx \rho(x)^2
\end{eqnarray*}

with $\psi_{R/L}$ being right/left moving fermi field, and $J$ and
$J_z$ are some average values of $J_{ij}$ and $J^z_{ij}$
respectively. As a reminder, the fixed point can be obtained by carrying
out a Wigner-Jordan tranformation and then linearizing the resulting
fermions around the Fermi points $\pm k_F$ (see Fig. 4). 

\begin{center}
\includegraphics[width=0.44 \linewidth]{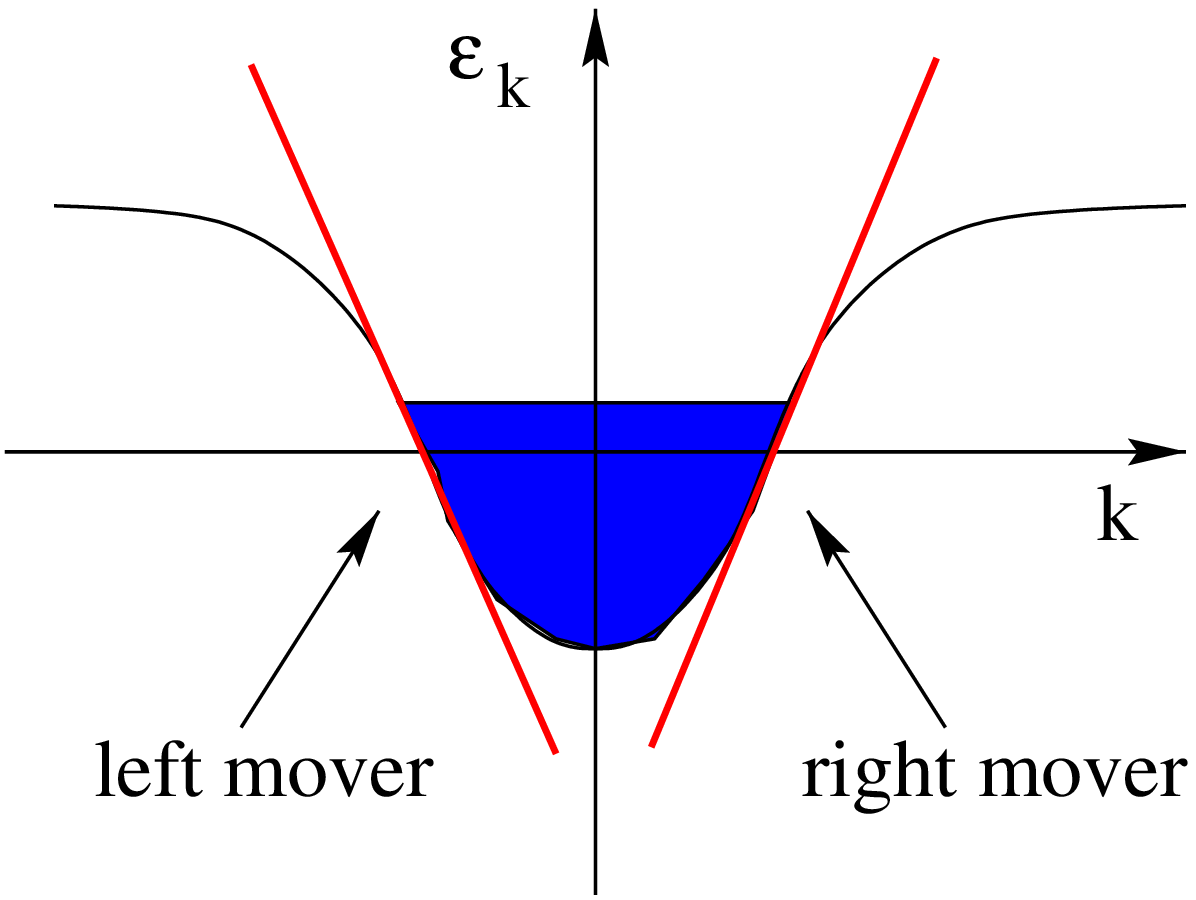}
\end{center}
\centerline{FIG. 4}
\bigskip

It will be convenient to change to bosonic variables in terms of which the 
fixed point hamiltonian takes the form (to leading order in $|J_z|/ J$),

\begin{eqnarray*}
 H_{LL}=v\int \frac{dx}{2 \pi}
\left( K (\pi\Pi)^2+\frac{1}{K} (\partial_x \phi)^2 \right) 
\end{eqnarray*}
where
\begin{eqnarray*}
v\approx\left(J+\frac{J_z}{\pi}\right)a\; , \quad K\approx\frac{1}{1+\frac{2J_z}{\pi J}}.
\end{eqnarray*}

We now consider the irrelevant operators around the fixed point. The
irrelevant operators come from various sources. We shall divide them
to Umklapp and non-umklapp operators. To the latter category belong
local terms coming from band curvature around $k_F$: for example $
\int \psi^\dagger_R \partial^2 \psi_R $.  However, the operators that
are important for transport are the Umklapp operators that reflect the
underlying lattice structure. Only they break translation invariance
and can degrade the currents. They have the following structure:

\begin{eqnarray*}
 H^{U} &=& \sum_{nm} H^U_{nm} \\
H^U_{nm} &=& g^U_{nm} \int \! dx [e^{i \Delta
k_{nm}x} \prod_{j=0}^{n} \psi_R^\dagger(x+j a) \psi_{L }(x+j a)
+h.c. ] \\
&=& \frac{ g^U_{nm}}{(2 \pi a)^n} \int dx 
[ e^{i \Delta k_{nm} x}
e^{ i  2 n \phi(x) }+ h.c.]
\end{eqnarray*}
where $ \Delta k_{nm}= n 2 k_F-m G$ is the momentum transfer
associated with the Umklapp process where $n$ particles are
transferred from one Fermi point to another while giving up to the
lattice $m$ units of lattice momentum $G= 2 \pi/a$. The particular values
of the couplings $g^U_{nm}$ depend on the couplings in the microscopic
model. These terms are irrelevant perturbations around the
fixed point not only by power counting but also because the $x$-dependent
exponentials in them suppress their contribution
exponentially. However they are the only source of dissipation.

We now consider the complete system consisting of an array of parallel
spin chains interacting with 3-dimensional acoustic phonons.  The 3-d
phonon system projected along the axis describing deformations of the
lattice {\em parallel} to the chains is described by
\begin{eqnarray*}
H_{p}^{3D}=  \int \frac{d^3x}{2\pi} 
\left[ (\pi P)^2 + 
\sum_\mu  v_{\mu}^2 ( \partial_\mu q)^2 \right],
\end{eqnarray*}
with $q$ the lattice deformation parallel to the spin chains direction, and
$P$ the conjugate momentum.

Integrating the corresponding propagator over the perpendicular
directions, we obtain the propagator along the chains:

\begin{eqnarray*}
\int d^2k_{\perp}\; \frac{1}{(\omega^2
+ \sum_\mu v_{\mu}^2 k_{\mu}^2)} &\sim& \ln [ (\omega^2 + v_p^2
k^2)/\Theta_{D}^2]. 
\end{eqnarray*}

In real space it takes the form $ 1/(x^2 + v_p^2 t^2)$ with $v_p$
being the slowest phonon velocity. This again is the propagator of a
$K=1$ Luttinger liquid describing the phonons, $H_p =v_p\int
\frac{dx}{2 \pi} \left( (\pi\Pi)^2+ (\partial_x q)^2 \right)$.  Again
we need to add to the combined phonon-spinon fixed point
$H^*=\sum_\alpha H^\alpha_p+\sum_\alpha H_{LL}^\alpha$ (the summation
$\alpha$ is over the spin chains) all the irrelevant operators. As
before they fall into two categories, the Umklapp and non Umklapp
operators.  The most important ones of the former one have the form,
\begin{eqnarray*}
H^{U,s-p}_{nm} = \frac{ g^{U,p}_{nm}}{(2 \pi a)^n} 
\int dx  [e^{i \Delta k_{nm} x}
e^{ i  2 n \phi } \partial_x q + h.c.]
\end{eqnarray*}
(where $\phi$ is the bosonic field in a particular chain),
while an example of the latter is $
H_{s,p}^{nonU} = \int (\partial\phi)^2\partial_x q $.

\bigskip

 We now turn to our main interest, computing the thermal conductivity
$\kappa_s(\omega,T)$ of the spin chains coupled to phonons. The thermal
conductivity can be expressed in terms of a heat current correlation
function,
\begin{eqnarray*}
\kappa(\w,T)=\langle J_Q, \; J_Q \rangle (\w,T) /\w
\end{eqnarray*}
where $J_Q$, the heat current expressed in bosonic variables, is $ J_Q=
-\sum_\alpha \int dx \, v^2\Pi_\alpha \partial_x \phi_\alpha -\int
d^3x\, v_p^2 P \partial_x q$ . The correlation is to be computed with respect to the
low-energy hamiltonian,
\begin{eqnarray*}
H_{low-E} = H^* + H^{U} + H^{nonU}. 
\end{eqnarray*}

As explained in the outline we need to identify 
the   (approximately) conserved ``charges'' of the low-E Hamiltonian, to find out whether they induce a slow decay of the heat current. We now show that the quantities,
\begin{eqnarray*}
J_s &=& v K \sum_\alpha \int dx \,
[\psi_{R\alpha}^\dagger \psi_{R\alpha}- \psi_{L\alpha}^\dagger \psi_{L\alpha} ]= v K \sum_\alpha\int dx \Pi_\alpha \\
P_T&=& -\; \; \; \sum_\alpha \int dx \Pi_\alpha \partial_x \phi_\alpha \; \; 
\; - \; \; \int d^3 x P \partial_x q 
\end{eqnarray*}
where, $J_s$ is the spin current and  $P_T$ the momentum operator, are 
 the  ``slow modes'' which in turn protect $J_Q$ rendering it slow too.
 
 Indeed, $J_s$ and $P_T$ commute with $H_{LL}$ and with $H^{nonU}$ and
 their conservation is violated only through $H^{U}$ with which they
 do not commute, thus inducing a slow current decay. 
More importantly, certain linear combinations of $J_s$ and $P_T$,
the ``pseudo-momenta''
\ba P_{nm} =
\frac{1}{2n}\Delta k_{ n,m} J_s + P_T \nonumber 
\ea 
decay even slower as they commute  with $H_{LL}+H^{nonU}+H^U_{nm}+H^{U,s-p}_{nm}$ and are therefore exactly conserved if only a {\em single} type of Umklapp with quantum numbers $n$ and $m$  is present. We note that
 the pseudo momenta can be written
as $P_{nm} = P_{lat} + \frac{m}{2n} G (N_R-N_L)$ with $P_{lat} =
\sum_k k c^{\dagger}_k c_k \approx k_F(N_R -N_L) +P_T$ being the
lattice momentum.  Unlike other (approximately) conserved quantities, they
 decay exponentially slowly with the temperature as their violation
requires processes away from the Fermi energy\cite{RA2}.

The heat current
 on the other hand does not commute with both $H^{nonU}$ and $H^{U}$
 and would therefore decay fast, but is protected by $J_s, P_T$ and their linear combinations since
 $\chi_{J_Q, J_s}, \chi_{J_Q, P_T} \neq 0$.

\bigskip

We proceed to discuss transport in the presence of several
approximately conserved - ``slow'' - variables: $J_1, J_2...J_N $. We
shall introduce the memory matrix formalism\cite{forster} which is very effective
under the circumstances as it allows the separation of the slow modes
($J_s, P_T, J_Q$, in our particular case) from the fast modes.

To set up the formalism one introduces a scalar product in the space of
 operators of the theory
\begin{eqnarray} 
\left(A(t)|B\right)&\equiv& \frac{1}{\beta} \int_0^\beta d\lambda
\left\langle A(t)^\dagger B(i \lambda) \right\rangle . \nonumber
\end{eqnarray}

In terms of this scalar product one can express the dynamic
correlation functions as follows,
\begin{eqnarray*} C_{AB} (\omega) &=&
\int_0^{\infty} dt e^{i \omega t} \left(A(t)|B\right)\\ &=& \left( A
\left| \frac{i}{ \omega -{\mathcal L}} \right| B \right)
\\ &=& \frac{iT}{ \omega} \int_0^{\infty} dt
e^{i \omega t} \left\langle [A(t),B]\right\rangle-\frac{(A|B)}{i
\omega}
\end{eqnarray*}

where the Liouville operator ${\mathcal L}$ is defined as ${\mathcal L}A = [H,A]$.
 
The matrix of conductivities (Kubo formula) is then,
\begin{eqnarray*}
\hat\sigma_{pq}(\w,T) =\frac{1}{T V} C_{J_p J_q} (\omega)
\end{eqnarray*}
$(p,q = 1 \cdots N)$.  In our case the thermal conductivity is
\begin{eqnarray*}
\kappa_s(\omega,T)=\frac{1}{T} \sigma_{QQ}(\omega,T). 
\end{eqnarray*}

However the dc conductivity has no good perturbative expansion: $
\sigma \sim 1/\Gamma$, with $\Gamma$ the decay rate of the current
is singular in perturbation theory.  Put in other words, if we
wish to compute the conductivity as a perturbative expansion of the
irrelevant operators around the fixed point, this would be an arduous
task in view of the fact that the conductivity computed from the fixed
point is infinite. Furthermore, even the perturbative expansion
  for $1/\sigma(\omega=0)$ turns out to be ill behaved in the presence
  of slow modes. While the short-time decay rate (see Fig. 2) of the current
is perturbative, the $\omega \to 0$ limit requires to take into account the presence of approximate conservation laws.

We seek therefore to compute a quantity which has a good perturbative
expansion, and to this purpose introduce $\hat{M}(\omega, T)$ - the
Memory Matrix, essentially the matrix of relaxation rates.

The matrix is defined as 

\begin{eqnarray*}\
\hat{M}_{pq}(\w) =\frac{1}{T}
\left(\partial_t J_p \left| {\mathcal Q} 
\frac{i}{\w-{\mathcal Q}{\mathcal L}{\mathcal Q}} {\mathcal Q} 
\right| \partial_t J_q \right)
\end{eqnarray*}

where ${\mathcal Q}$ is the projection away from slow modes 
\begin{eqnarray*}
{\mathcal Q}=1-\sum_{pq} |J_p) \frac{1}{T} (\hat{\chi}^{-1})_{pq} (J_q|.
\end{eqnarray*}

 In terms of the memory matrix the conductivity matrix is,
\begin{eqnarray*}
 \hat\sigma(\omega,T)=\hat{\chi}(T) \left(\hat{M}(\omega,T)- i \w
 \hat{\chi}(T) \right)^{-1} \hat{\chi}(T)
\end{eqnarray*}

with $\hat{\chi}$ the susceptibility matrix,
\begin{eqnarray*}
 \hat{\chi}_{pq}= \frac{1}{T V} (J_p|J_q).
\end{eqnarray*}

\bigskip

Applying the formalism in our case we find that the memory matrix is a
sum over the Umklapp processes ($nm$), given - to leading order in
$g^U_{mn}$ - by,
\begin{eqnarray*}\label{MM}
\hat{M} = \frac{1}{T} \left[ \sum_{nm} (\hat{M}_{nm}+\hat{M}_{nm,s-p})\right] 
\end{eqnarray*}
where (the matrix indices $p,q$ take the values  $s,T,Q$)
\begin{eqnarray*}
 M_{nm}^{pq}&\equiv& \frac{
\langle F^p;F^q \rangle^0_\w-
\langle F^p;F^q \rangle^0_{\w=0}}{i \w}\; ,  \\
 M_{nm,s-p}^{pq}&\equiv& \frac{
\langle F_{s-p}^p;F_{s-p}^q \rangle^0_\w-
\langle F_{s-p}^p;F_{s-p}^q \rangle^0_{\w=0}}{i \w}\; ;
\end{eqnarray*}

here $F^p=i[J_p,H^U]$, $F_{s-p}^p=i[J_p,H^{U,s-p}]$ and $\langle
 F^p;F^q\rangle^0_{\w}$ is the retarded correlation function
 calculated with respect to $H^*$. As the perturbative expansion is in
 irrelavant operators with respect to the fixed point the expansion is
 expected to be rapidly converging at low temperatures if slow and
 fast time scales are well seperated and all of the slowest modes have
 been taken into account.

Carrying out the computation of the various correlation functions (see
Ref.~[\onlinecite{SAR}] for details) we have:
\begin{eqnarray*}
\kappa_s (T) & \approx & v^2 T^3 \left[(\hat{M}^{-1})_{TT}
+2(\hat{M}^{-1})_{QT}+(\hat{M}^{-1})_{QQ}\right]
\end{eqnarray*}
with the typical matrix elements,
\begin{eqnarray*}
{M}_{nm}^{pq}&\sim& (\Delta k_{nm})^{(n^2 K-2)} e^{-v \Delta k_{nm}/2T}\\
{M}_{nm,s-p}^{pq}&\sim& T^{(2 n^2 K - 1)} e^{-v_p \Delta k_{nm}/2T}.
\end{eqnarray*}

Note that the spinon processes decay exponentially fast, with the
exponent $-v \Delta k_{nm}/2T$, while the spinon-phonon exponents
contain the much slower phonon velocity $v_p\ll v$,
leading to a much slower decay with the exponent $-v_p \Delta k_{nm}/2T$. 
The latter will therefore clearly determine the thermal conductivity.

\bigskip

But which of the scattering processes $(n,m)$ will dominate?

At low-T exponential factor prevails hence the smallest $
\Delta k_{nm}$. At this point our discussion must distinguish between
commensurate and incommensurate magnetization, or in fermionic language
commensurate and incommensurate filling. The magnetization can be 
in principle tuned by varying an external magnetic field $h$.

Close to commensurate filling $k_F \approx G \frac{m_0}{2n_0}$, and
then the dominant processes appear to be $ H^U_{n_0 m_0}$ where $
\Delta k_{n_0,m_o} \approx 0$.  However, because of the conservation
laws discussed earlier, it is the next leading term $H^U_{n_1 m_1}$
with $ \Delta k_{n_1,m_1}= \pm G/n_0$ which determines the decay rate.
Technically, this arises as one of the eigenvalues of the matrix
  $\hat{M}$ is not affected by $ H^U_{n_0 m_0}$ (as $[P_{n_0
    m_0},H^U_{n_0 m_0}]=0$) but determined by $H^U_{n_1 m_1}$. This
  smallest eigenvalue will then determine the size of $\hat{M}^{-1}$
  and therefore the heat conductivity. The particular case of half
filling is discussed in detail below.

On the other hand,
 at a typical incommensurate filling it will depend on the temperature
which processes are dominant and subdominant
and we need to sum over all terms (do saddle-point approximation with
 respect to  $n_1$) and find,
\begin{eqnarray*}
 \kappa_{\text{typical}}\sim \exp[c (\beta v G)^{2/3}]\; . 
\end{eqnarray*}
with  $c$ a constant of order 1.
\bigskip

We now turn to the experiment by Sologubenko {\it et al.} discussed
earlier. The experiment was carried out at $h=0$ corresponding to half
filling, $k_F= \pi/2a = G/4$. Therefore $ \Delta k_{21} =0$ (n=2,
m=1). But we need at least two Umplapp terms and the next smallest is
$\Delta k = G/2, (n=1,m=0)$. Recall also that as $ v_p \ll v $, it follows that
$\hat{M}_{nm,s-p} \gg \hat{M}_{nm}$.
Hence, the dominant contribution to the thermal conductivity comes
from, $(\hat{M}^{-1})_{TT} \approx 1/ M_{n=1, s-p}^{TT}(G/2,T) $ and
we have

\begin{eqnarray*}
 \kappa(T) \approx \kappa_0(T/T^*)^{2(1-K)} e^{T^*/T}
\end{eqnarray*}
with
\begin{eqnarray*}
 T^* = v_p G/4\; . 
\end{eqnarray*}

We find therefore that the second strongest rate wins (cf. the
expression for $\hat{M}$): it is determined by $v_p$ via a phonon
process and is characterized by the momentum $G/2$.  The $G/2$
 transfer momentum characterizes
 the dominant spinon - phonon Umklapp process, and clearly
distiguishes it from pure phonon Umklapp processes characterized by
momentum transfer $G$. We expect therefore that in the pure phonon
thermal conductivity (axes $a$ and $c$ in Fig. 3) the scale $2 T^*$
would appear.

\bigskip

To compare our findings with the experimental data we need to express
our expression in terms of $\Theta_D$. Assuming an isotropic phonon dispersion
one has:  $\Theta_D \approx
  v_p(6\pi^2/a^3)^{1/3}\approx 0.6 \, v_p G$. 

Therefore: 
\begin{eqnarray*}
T^\ast &\approx&  0.4\Theta_D, \text{(theory)}\\
T^\ast &\approx&  0.42\Theta_D, \text{(experiment)}.
\end{eqnarray*}
Taking into account possible ambiguities in the fits to the
  experiments and that the phonon dispersion is probably not
  completely isotropic, part of this excellent agreement may be
  accidential. But further confirmation of our theory comes from the
  observation that the ratio of slopes of the spinon contribution
  compared to the pure phonon contribution (on a semilogarithmic graph
  of $\kappa$ vs. $1/T$ - see Fig. 5\cite{ssolo} ) is approximately
  1:2 as discussed earlier.

\begin{center}
\includegraphics[width=0.70 \linewidth]{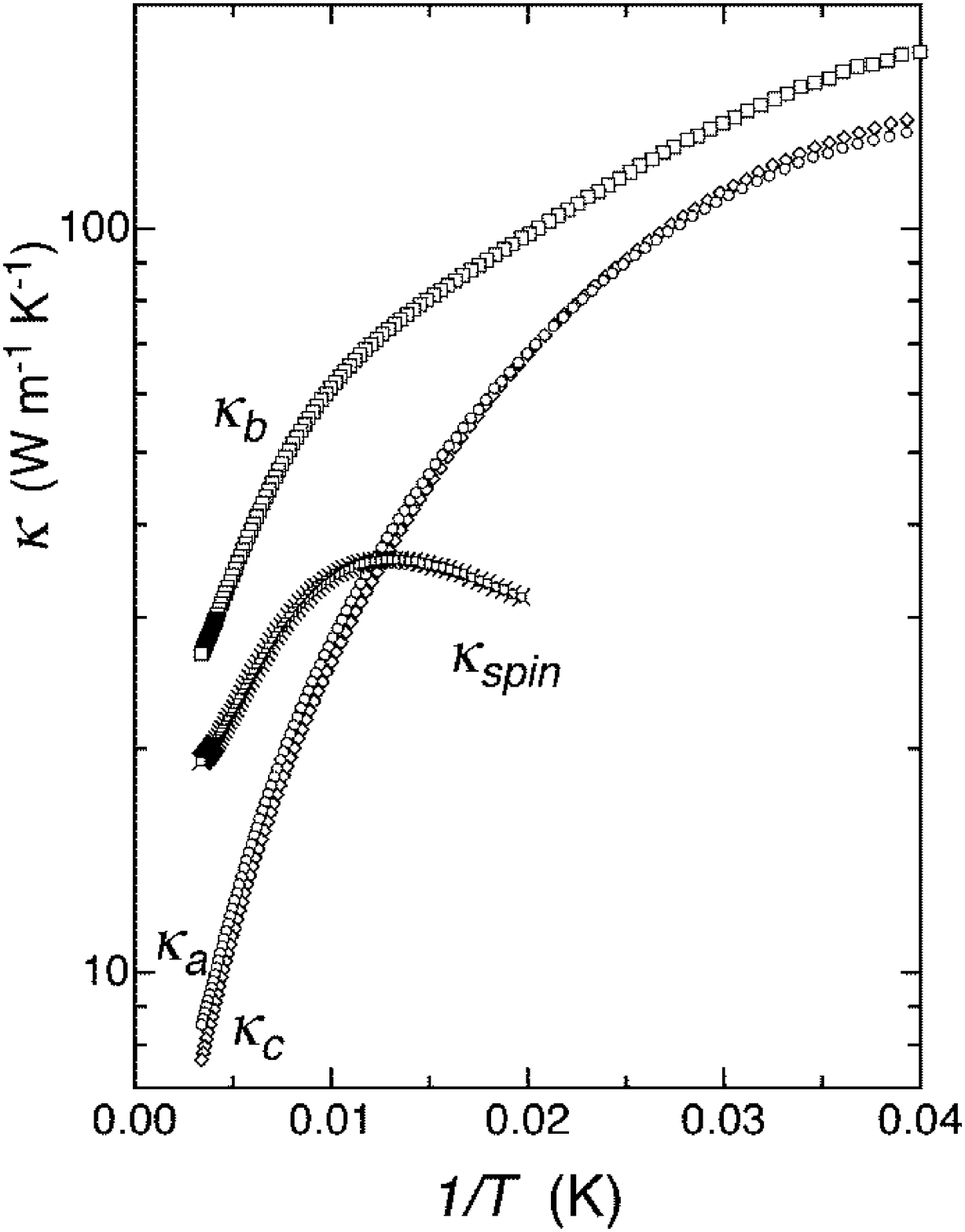}
\end{center}
\centerline{FIG. 5}
\bigskip

We may also consider the effect of a magnetic field on the thermal
conductivity. (To observe it experimentally one needs a material with
much smaller spin energy scales $J$ than SrCuO). The magnetic field 
modifies the value of $k_F$ according to
\begin{equation}
k_F=\frac{\pi}{2 a} (1+M)\approx\frac{\pi}{2 a} (1+h/(\pi J)). \nonumber
\end{equation}

As the field  $h$ is varied the system passes through 
 commensurate fillings,  $\Delta k_{nm}= n 2 k_F-m G =0 $
and  incommensurate fillings,  $\Delta k_{nm}= n 2 k_F-m G \neq 0 $.
Thus different Umklapp operators become effective leading to a fractal-like
 dependence on  $M\approx h/\pi J$ (see Fig. 6). 

\bigskip
\begin{center}
\includegraphics[width=0.95 \linewidth,clip=]{fractal.eps}
\end{center}
\centerline{FIG. 6}
We expect some dips to be experimentally observable.

\bigskip

In conclusion,

\begin{itemize}

\item Transport  is strongly influenced by conserved ``charges'': low
energy processes cannot relax heat current.

\item  Exponents are determined by slowest  mode in the system: typically phonons. 

\item Memory Matrix approach, separating slow and fast modes, allows 
controllable calculations.

\item Calculations fit experiments

\item Interesting predictions on the magnetization dependence of the heat transport 
have been made. 

\end{itemize}

\end{document}